\documentclass[]{article}
\usepackage{lineno}
\usepackage{authblk}
\usepackage{graphicx}
\usepackage{listings}
\usepackage{xcolor}
\usepackage{bm}
\usepackage{algorithm}
\usepackage{algpseudocode}
\usepackage{caption}
\usepackage{subcaption}
\usepackage{amsmath}
\usepackage{siunitx}
\usepackage{hyperref}
\usepackage{sectionbreak}
\usepackage{stackengine}

\definecolor{codegreen}{rgb}{0,0.6,0}
\definecolor{codegray}{rgb}{0.5,0.5,0.5}
\definecolor{codepurple}{rgb}{0.58,0,0.82}
\definecolor{backcolour}{rgb}{0.95,0.95,0.92}

\lstdefinestyle{mystyle}{
    backgroundcolor=\color{backcolour},   
    commentstyle=\color{codegreen},
    keywordstyle=\color{magenta},
    numberstyle=\tiny\color{codegray},
    stringstyle=\color{codepurple},
    basicstyle=\ttfamily\footnotesize,
    breakatwhitespace=false,         
    breaklines=true,                 
    captionpos=b,                    
    keepspaces=true,                 
    numbers=left,                    
    numbersep=5pt,                  
    showspaces=false,                
    showstringspaces=false,
    showtabs=false,                  
    tabsize=2
}

\title{SOniCS: Develop intuition on biomechanical systems through interactive error controlled simulations}
\author[1]{Arnaud Mazier*}
\author[2]{Sidaty El Hadramy*}
\author[3]{Jean-Nicolas Brunet}
\author[1]{Jack S. Hale}
\author[2]{Stéphane Cotin}
\author[1]{Stéphane P.A. Bordas}
\affil[1]{Institute of Computational Engineering and Sciences, Department of Engineering, Université du Luxembourg, 6, Avenue de la Fonte, 4364, Esch-sur-Alzette, Luxembourg}
\affil[2]{MIMESIS team, Inria, 1 Place de l’Hopital, 67000 Strasbourg, France}
\affil[3]{jnbrunet@pm.me}
\affil[*]{Arnaud Mazier and Sidaty El Hadramy contributed equally to this work}
\begin{document}

\maketitle

\begin{abstract}
This new approach allows the user to experiment with model choices easily and quickly without requiring in-depth expertise, as constitutive models can be modified by one line of code only. This ease in building new models makes SOniCS ideal to develop surrogate, reduced order models and to train machine learning algorithms for uncertainty quantification or to enable patient-specific simulations. SOniCS is thus not only a tool that facilitates the development of surgical training simulations but also, and perhaps more importantly, paves the way to increase the intuition of users or otherwise non-intuitive behaviors of (bio)mechanical systems. The plugin uses new developments of the FEniCSx project enabling automatic generation with FFCx of finite element tensors such as the local residual vector and Jacobian matrix. We validate our approach with numerical simulations such as manufactured solutions, cantilever beams, and benchmarks provided by FEBio. We reach machine precision accuracy and demonstrate the use of the plugin for a real-time haptic simulation involving a surgical tool controlled by the user in contact with a hyperelastic liver. We include complete examples showing the use of our plugin for simulations involving Saint Venant-Kirchhoff, Neo-Hookean, Mooney-Rivlin, and Holzapfel Ogden anisotropic models as supplementary material.
\end{abstract}

\section{Introduction}
SOniCS is an ideal tool to develop more intuitive understanding of biomechanics systems. It enables the development of accelerated surrogate models for parameter identification and uncertainty quantification, model selection through machine learning~\cite{Deshpande2022} or model order reduction~\cite{Goury2016}. It also enables fast and hypothesis testing possibilities to non-experts. Designing efficient finite element (FE) simulation software is a challenging task. Indeed, as FEM is a vast field, numerous pieces of software emerged to fill different gaps. For instance, commercial software such as Abaqus~\cite{Abaqus2009} or Ansys~\cite{Ansys1985} focus on proposing a user-friendly GUI (Graphical User Interface) guiding the user from pre-processing to post-processing. This has the advantage of allowing users to perform complex simulations with a relatively basic theoretical knowledge of FE. Meanwhile, other pieces of software focused on specific domains such as Gmsh~\cite{gmsh} for FE meshing, Paraview~\cite{Paraview} for FE visualization, OpenFoam~\cite{Openfoam} for CFD (Computational Fluid Dynamic) simulations, OpenXFEM~\cite{Bordas2006} for extended finite elements~\cite{Jansari2019}, collocation methods~\cite{Jacquemin2021}, meshfree methods~\cite{Nguyen2008}, multiscale problems~\cite{Talebi2013}, or material point methods (MPM)~\cite{Bordas2017}. This historical perspective explains the countless FE solvers which makes an exhaustive state-of-the-art review quasi-impossible. Before selecting one piece of FE software, an user should consider the benefits and disadvantages associated with each (i.e., meshing, parallel support, solvers, coding language, and visualization). However, generally, simplicity of use and the ability to easily test modelling hypotheses appears like the most important considerations in selecting such a computational tool.

In the medical simulation context, several aspects have to be considered.
\begin{itemize}
	\item The material model complexity. Conversely to engineering materials such as steel or copper, the mechanical properties of living organs were only recently quantified~\cite{Chester1934, Payan2017} (early 90s against 1700 for copper) and show immense variability~\cite{Mihai2017}. Various models have been proposed, e.g., anisotropic~\cite{Zhou1997, Picinbono2003, Flynn2011, Boyer2013, Elouneg2021}, hyperelastic~\cite{Martins2006, Mikhail2001, Mihai2015, Chagnon2015, Milad2021}, viscoelastic~\cite{Ehlers2001, Haj-Ali2004, Marchesseau2010, Urcun2021}, or poroelastic~\cite{Simon1992, Cowin1999, Stokes2010, Scott2021, Barrera2021, Bulle2021, Lavigne2022} to accurately depict their complex mechanical behaviors. Indeed, predicting the deformations of bio-materials can only be achieved through complex material models (sometimes even multi-scale), which are rarely implemented in commercial software. One major difficulty of this implementation remains the differentiation of highly nonlinear equations. Indeed, predicting the deformations of such materials can only be achieved through complex material models, rarely implemented in commercial software. One major difficulty of this implementation remains the derivation of highly nonlinear equations. For instance, hyperelasticity equations can be written as a minimization of a tensorial function. In most cases, this minimization is solved using gradient-descent algorithms requiring the first and second derivatives of the functional. Therefore, obtaining such high-order nonlinear derivatives is not straightforward and can be prone to manual errors.
	
	\item The complexity of the simulation. In addition to the complexity of the material model, the simulation setup itself can be problematic. For instance, the material parameters are patient-specific and require data-driven or inference methods~\cite{Han2011, Urcun2021-2}. The simulations can also (partially) involve unknown boundary conditions~\cite{Tagliabue2021}, contact with other organs~\cite{Courtecuisse2014, Miguez2015}, or surgical tools~\cite{Cotin1999, Lim2006}. The problem can also require multi-physics models (such as FSI~\cite{Borazjani2013, Bianchi2017} (Fluid-Structure Interactions)) or incompressibility~\cite{Weiss1996, Mazier2022}, where classical displacement-based finite elements are prone to locking.
	
	\item The error control and uncertainty of the solution. When dealing with biomechanical simulations, several uncertainties always arise from the material parameters, loads, geometry, or boundary conditions. This is mainly due to the difficulty in estimating the mechanical properties through ex-vivo methods or the topology of biological tissues using medical imaging. Similarly, error control and mesh adaptivity are necessary to ensure homogeneous convergence of the solution over the domain and that the mesh is optimal given a quantity of interest~\cite{Allard2012}. Therefore, quantifying the uncertainty or controlling the error on quantities of interests often requires making a very large number of simulations~\cite{Rappel2019, Rappel2019-2} which is incompatible with surgical timing~\cite{Hauseux2017, Hauseux2018}. Meanwhile, those approaches could be functional in clinical settings by using accelerated simulations~\cite{Bui2018} to build surrogate models or/and machine learning models for faster solutions of those highly non-linear parametric problems. 
	
	\item The real-time aspect. In addition to the previous point, for clinical environments, the running time of the numerical simulation is crucial. When performing an operation, the surgeon cannot, in general, spend minutes waiting for the model predictions. Eventually, this aspect is also applicable to artificial intelligence. Indeed, in order to build an efficient machine learning model, a significant amount of data is necessary. Using numerical simulations to create synthetic data is now standard and directly dependent on the simulation time~\cite{odot2022, Deshpande2022}. Consequently, the run-time of the simulation can be considered a principal feature of biomechanics simulations, and indeed, of any non-intuitive non-linear problems subject to significant uncertainties in loading, boundary and initial conditions, and parameters. Nowadays, gaming engines such as Unity3D~\cite{Haas2014} or Unreal Engine~\cite{Unrealengine} offer real-time animation where physics-based algorithms can be included~\cite{Comas2008, Verschoor2018, Turini2019}.
	
	\item The interaction with the user. In a surgical simulation setting, external variables can impact the simulation during the execution. For example, the exact movement of the surgeon's tool influences the simulation during run-time. Thus, the parameters of the simulation must be tuned, "live", to integrate interactions between the user and the simulation~\cite{Niroomandi2013, Wu2014}.
	
	\item The visual rendering. Depending on the research field, visualization can play a critical role in the understanding of the results. For instance, in the computer graphics community, photo-realistic visualization is one of the main objectives~\cite{Gilles2011, Malgat2015}. Contrastingly, in the mechanical engineering culture, visualization is a manner of extracting and understanding quantitatively a solution or data set (i.e., stress or displacement fields). For medical simulations, an optimal solution must combine both the accuracy of the results and photo-realistic rendering reflecting the clinical ground truth all within clinical time frames~\cite{Guo2021}.

\end{itemize}

According to the state-of-the-art, SOFA~\cite{faure:hal-00681539} (Simulation Open Framework Architecture) appears to be a suitable compromise. Indeed, SOFA employs efficient rendering while providing the possibility to interact in real-time with the running simulation. One can note that real-time computing is only possible depending on the complexity of the problem. Indeed, using excessive numbers of degrees of freedom (DOFs) or solving a highly nonlinear problem cannot result in a real-time simulation (without using model order reduction~\cite{ Chinesta2014, Goury2016, Goury2018} or machine learning~\cite{Deshpande2022}). SOFA can also manage complex simulations through an efficient implementation of contact, for example, or enabling multi-physics coupling. Therefore, only a few material models and elements are available. A similar issue is observed for material models where numerous implementations only focused on a 3-dimensional isotropic behavior. Therefore, coding a new material model or element in SOFA requires advanced C++ skills that may discourage individuals from using the software.

To alleviate the problem of complex material models, FEniCS~\cite{AlnaesEtal2015} seems like an appropriate solution. Indeed, FEniCS may not possess all of SOFA's features, but definitely overcomes SOFA's capabilities for material model complexity. With FEniCS, the user can generate a large number of material models, regardless of the element's geometry or interpolation. Plus, it authorizes an export of the pertinent finite element tensors in C code to be efficiently plugged into SOFA. The benefits of the synergy are considerable. By using SOFA's interactivity and real-time features, the user can easily prototype a real-time simulation. Indeed, by modifying, in "live", various boundary conditions, geometries, or topologies, the user can effortlessly and rapidly verify modeling hypotheses of a specific problem. Combined with the specificities of FEniCS, the user can additionally smoothly prototype complex material models for modeling elaborate scenarios. Such feature has already been used by coupling FEniCS and Acegen but with different objectives~\cite{fenics2021-lengiewicz}. To the authors' knowledge, this paper is the first to use FEniCS code generation capabilities for such an endeavor and is the first coupling between FEniCS and SOFA. 

This paper has the following outline. We will first briefly introduce SOFA and FEniCS, highlighting the relative advantages and design choices in each. Section \ref{sec:sonics} will detail the plugin functionalities and a short tutorial for importing a Saint Venant-Kirchhoff model from FEniCS in SOFA. Then, section \ref{sec:analysis} will focus on confirming our implementation for various numerical tests. We will use a manufactured solution in \ref{sec:manufactured} as validation and compare our solutions for a cantilever beam problem with SOFA in \ref{section:beam}. The last test consists in implementing a new material model (Mooney-Rivlin) in SOFA and benchmarking it with FEBio~\cite{febio} in \ref{sec:febio}. Finally, in the last section \ref{sec:haptic}, we will use our plugin in a complex haptic simulation that cannot be implemented in FEniCS, using a custom material model inexistent in SOFA.
\sectionbreakmark{\rule{10em}{3pt}}
\sectionbreak

\subsection{SOFA}
SOFA was created in $2007$ by a joint effort from Inria, CNRS, USTL, UJF, and MGH. Such piece of software aims to provide an efficient framework dedicated to research, prototyping, and the development of physics-based simulations. It is an open-source library distributed under the LGPL license, hosted on GitHub at \url{https://github.com/sofa-framework/sofa}, and developed by an international community. SOFA is modular. Users can create public or private plugins to include additional features.

SOFA is a C++ library, including Python wrappers for a user-friendly prototyping interface. It was originally designed for deformable solid mechanics but has been extended to various domains such as robotics, registration, fluid simulations, model-order reduction, and haptic simulations~\cite{Duriez2006, Duriez2013}. SOFA exhibits many attractive features, but among them, the combination of multi-model representations and mappings differentiate it from other software. 
\smallbreak
\noindent \textbf{Multi-model representation}: Most classical FE software uses an identical discretization for the whole model. Consequently, if one user wants to refine the mesh along a contact surface, the FE mesh will undergo the same refinement in the contact region. It can induce slow simulations for solving the FE system while the user was initially only interested in the contact part. Conversely, utilizing a multi-model representation approach, users can split the principal model into three distinct sub-models: deformation, collision, and visual. Thus, the user can decide to have high fidelity deformations with flawed contact detection while maintaining a fine rendering, or vice-versa. Similarly, an object can be made of several deformation models. For example, one can model a muscle by the interaction of FE 3D tetrahedra for the volume and 1D beams for modeling ligaments, using 2 different solvers.
\smallbreak
\noindent \textbf{Mappings}: In SOFA, the "mappings" are responsible for the communication between the different models. The models have parent-child relationships constructing a hierarchy (and a DOF hierarchy by extension). It enables propagating the positions, velocities, accelerations, and forces across the different models. For example, if the contact model calculates a force, it is mapped on the deformation model that will communicate back the computed displacement.

Finally, by combining the multi-model representation with the mappings, SOFA can build complex real-time simulations with high fidelity rendering. In addition to a scenegraph structure and visitors (responsible for going through the model hierarchy) implementation, it can account for interactivity with the users.

Despite the advantages provided by SOFA, some drawbacks have to be acknowledged. First, in terms of solid mechanics simulations, only a few elements and material models are available. Indeed, SOFA only proposes Lagrange linear elements, and the geometries are limited to segments, triangles, quadrangles, tetrahedra, and hexahedra. The following material models are coded: Boyce and Aruda~\cite{Arruda1993}, Costa~\cite{Costa2001}, isotropic and anisotropic Hookean, Mooney Rivlin (2 invariants)~\cite{Mooney1940}, classical and stabilized Neo-Hookean, Ogden~\cite{Ogden1972}, Saint Venant-Kirchhoff, and Veronda Westman~\cite{Veronda1970}. Despite a reasonable number of mechanical models, a few of them are actually implemented for each element type. Secondly, the benefits provided by the mappings can also turn out to be a disadvantage when it comes to implementation. Indeed, the structure of the mappings is usually complex for unexperimented C++ users, and the mechanical tensors such as the Cauchy-Green or Piola-Kirchhoff are rarely computed. The two previous drawbacks are associated to the same flaw: the strong coupling between the material models and the topology of the element assumed within SOFA's architecture. This coupling implies that changing an element's topology or interpolation will involve a new mapping or the rewriting of the material model, even in the case of a similar material model.

\subsection{The modular mechanics plugin (Caribou)}
The initial goal of the plugin (called Caribou at the time of writing, \url{https://github.com/mimesis-inria/caribou}) was to quickly implement new shape functions and their derivatives for different Immersed-Boundary and meshless domain discretization while keeping the compatibility with the existing SOFA surgical simulations~\cite{brunet2020}. Besides, the plugin enabled to effortlessly implement different volumetric quadrature schemes and several hyperelastic material models. Hence, the software design had to be generic enough to combine all the previous requirements. It also had to be efficient enough to avoid the creation of a bottleneck that would prevent the biomechanical model from meeting its computational speed requirement. Hence, the plugin was made as an extension to SOFA, bringing a redesigned software architecture. 

In the plugin, the authors implemented a compile-time polymorphism design using generic C++ template programming. The idea is to write the code as close as possible to equations found in traditional FE books. Then, the C++ compiler optimizes the set of operations executed during the simulation while keeping an object-oriented code. In this design, the "Element" concept was created as a generic computational class that would be inherited by all element types. Similarly to OpenXFEM++~\cite{Bordas2006}, it provides a flexible implementation to add interpolation and quadrature numerical procedures quickly. Since standard isoparametric elements have a number of nodes, quadrature points, and shape functions already known at compile-time, most modern compilers will be able to aggressively inline the code to optimize the computation.

Finally, the plugin allows the creation of additional material models by simply defining three methods per material: the strain energy density function, the second Piola-Kirchhoff stress tensor function, and its derivative functions. These three functions are evaluated at a given integration point automatically provided by the plugin. This design delivers an undeniable advantage: writing a new material model is now independent of the topology and integration scheme. However, it comes with a non-negligible cost. The author of the new material model has to manually differentiate the strain energy twice and write it in C++. This manual intervention is error prone and can quickly become a substantial drawback for complex materials.

\subsection{FEniCS}
The FEniCS Project (FEniCS) \cite{AlnaesEtal2015} is a collection of tools for
the automated solution of partial differential equations using the finite
element method. Like SOFA, the FEniCS components are distributed
under open-source licenses (LGPL v3 or later, and MIT) and development is
hosted on GitHub at \url{https://github.com/fenics}.

A distinguishing feature of FEniCS is the ability to allow the user to write
variational of weak formulations of finite element methods in a high-level
Python-based domain specific language (DSL), the Unified Form Language
(UFL)~\cite{alnaes_unified_2014}. Subsequently, that high-level description can
be compiled/transformed using the FEniCS Form Compiler
(FFC)~\cite{kirby_compiler_2006} into low-level and high-performance kernels.
These kernels can calculate the corresponding local finite element tensor for a given
cell in the mesh. UFL is also used by other finite element solvers with
independently developed automatic code generation capabilities, notably
Firedrake~\cite{rathgeber_firedrake_2016} and Dune~\cite{bastian_dune_2021}.
Compared with SOFA, FEniCS is limited in scope; its primary focus is the
specification and solution of partial differential equations via the finite
element method, leaving difficult problems like mesh generation,
post-processing and visualisation to leading third-party packages such as
Gmsh~\cite{gmsh} and Paraview~\cite{Paraview}.

In the context of implementing finite element models of hyperelastic materials,
this automatic approach has a number of advantages over the traditional route
used by most finite element codes (including, to some extent, SOFA);
differentiating analytical expressions for the residual (first derivative) and
Jacobian (second derivative) of the energy functional for the hyperelastic
model, picking a suitable finite element basis and then hand-coding the
corresponding finite element kernels in a low-level language (C, Fortran, C++)
for performance. Specifically:
\begin{enumerate}
	\item The symbolic residual and Jacobian can be derived automatically
	using the symbolic differentiation capabilities of
	UFL. By contrast taking these derivatives by hand can be tedious and error-prone.
	\item The compilation of the UFL description of the problem by FFC into the
	associated low level kernels is entirely automated. Again, this step is
	often time consuming and difficult to perform manually.
	\item Because of the high-level description of the problem it is possible to
	experiment quickly with different concrete finite element
	formulations (material models, basis functions, element
	topology etc.)\ without manually modifying low-level kernel
	code.
\end{enumerate}
The potential of this high-level approach for solid mechanics were recognised
early on in the development of FEniCS, with two chapters in the FEniCS
Book~\cite{olgaard_applications_2012,narayanan_computational_2012} promoting
this direction. Since then, FEniCS has been used in a large number of
publications on the topic of hyperelastic large-deformation elasticity e.g.~\cite{Baroli2012, Phunpeng2015, Weis2017, Nguyen2020, Patte2022}.

Recently the FEniCS Project has undergone a major redevelopment, resulting in
the new FEniCSx components; DOLFINx (the finite element problem solving
environment, replacing DOLFIN), FFCx (the FEniCSx Form Compiler, replacing FFC)
and Basix~\cite{BasixJoss} (a finite element basis function tabulator,
replacing FIAT~\cite{kirby_algorithm_2004}). UFL is largely unchanged from
the version used in the old FEniCS components and Firedrake.

In this work we do not use DOLFINx. DOLFINx contains the basic finite element data structures
and algorithms (e.g. meshes, function spaces, assembly, interfacing with linear
algebra data structures in e.g. PETSc~\cite{}) and therefore there is a
significant overlap with the functionality already available in SOFA. Directly
interfacing DOLFINx and SOFA at the Application Programming Interface (API)
level would be a significant technical challenge due to the substantial
differences in their internal data structures. Dedicated weak coupling
libraries such as PreCICE~\cite{rodenberg_fenicsprecice_2021} could be an interesting alternative to API
coupling, but it is not a path that we explore in this work.

Instead, the approach taken by SOniCS is to only use UFL and FFCx (which in
turn depends on Basix) to convert the high-level description of the finite
element problem into low-level C code, which are then called using SOFA's existing C++ finite element data structures and algorithms. Compared with
coupling DOLFINx and SOFA directly, our approach creates a relatively light
compile-time coupling between FEniCS (specifically, the generated C code) and
SOFA (a large complex C++ code with many dependencies). Consequently, no
additional runtime dependencies required for SOFA. This methodology will be
familiar to users of the DOLFINx C++ inteface where C finite element kernels
are generated in a first step using UFL and FFCx and are then integrated into the
DOLFINx solver in a second step through a standard compile/include/link
approach. Without going into excessive detail, two changes in the redeveloped
FEniCSx components have made SOniCS significantly easier to realise:
\begin{enumerate}
	\item Basix and FFCx have full support for Serendipity finite elements of
	arbitrary polynomial order following the construction of Arnold and
	Awanou~\cite{arnold_serendipity_2011}. Serendipity elements are
	used in SOFA and there was a desire to continue
	supporting Serendipity basis function due to their lower number
	of degrees of freedom per cell and generally lower number of
	local computations compared with standard tensor-product
	Lagrange elements. We remark that despite the widespread use of
	Serendipity elements in many solvers, they can only obtain
	optimal order convergence on affinely-mapped meshes, see
	e.g.~\cite{arbogast_direct_2022} for more details.
	\item FFCx outputs C99 compliant code according to the UFCx interface, which is
	specified as a C header file included with FFCx. This is in contrast with
	FFC, which outputs C++03 compliant code conforming to an interface specified with a C++
	header file. This switch makes it significantly easier to call
	FFCx generated kernels from libraries with a C Foreign Function
	Interface (FFI) such as Python and Julia, or any language
	which can easily call functions with a C ABI (e.g. Fortran).
	Although SOFA is a C++ libraries and could certainly call C++ generated
	kernels, the C interface is simpler to use, consisting only of structs containing
	basic native data types and functions. 
\end{enumerate}

\section{SOniCS}
\label{sec:sonics}
\begin{figure}[htp]
	\centering
	\includegraphics[width=\linewidth]{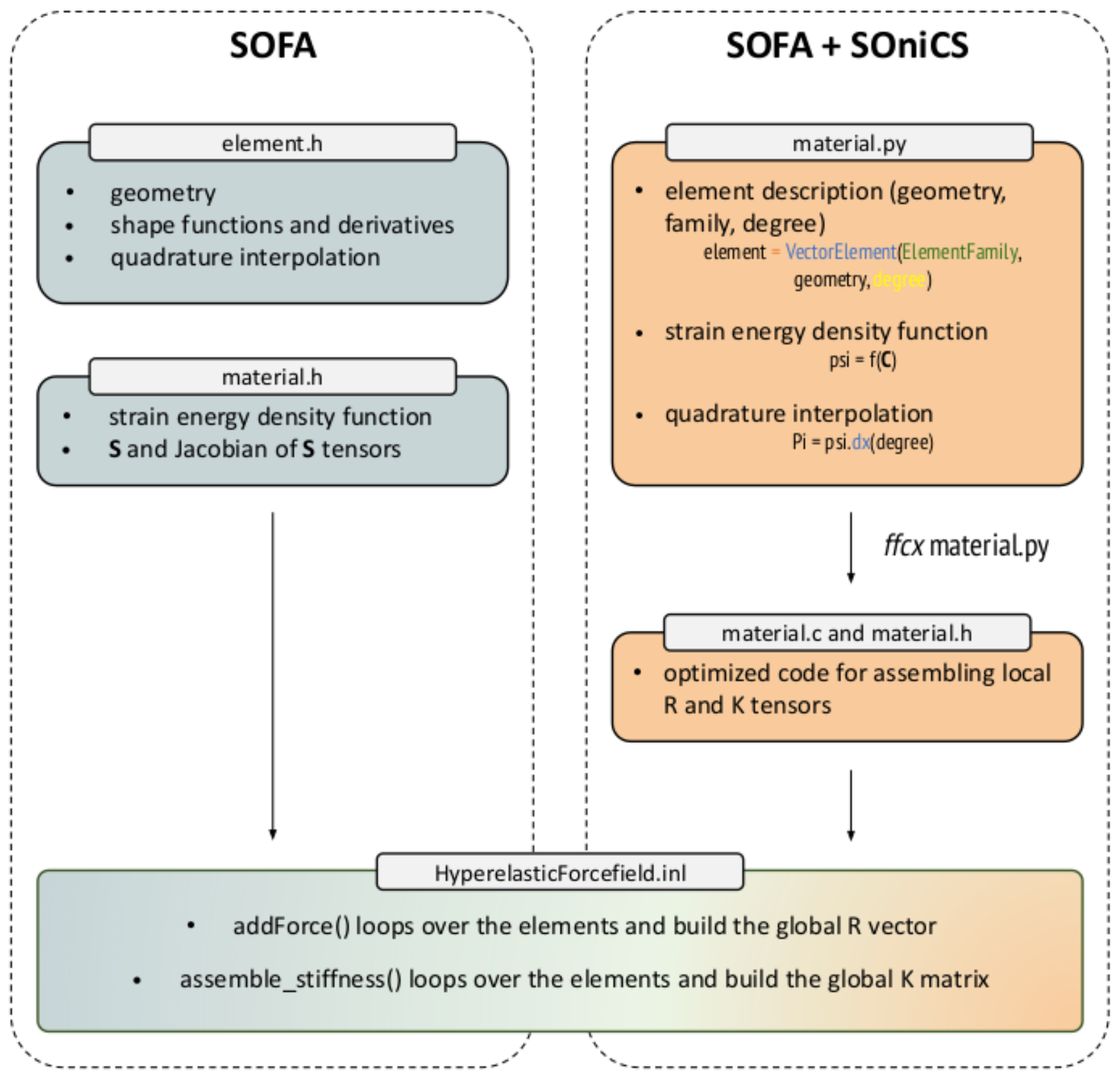}
	\caption{Description of the SOniCS pipeline (on the right) and differences with SOFA(on the left). In SOFA, each element has to be defined, embedding its geometry, shape functions (including derivatives), and the quadrature scheme and degree. In SOniCS, it has been replaced by two Python lines of code for describing the element and its quadrature. The same benefit goes for the material model description. In SOFA, each material has to be created in a separated file stating its strain energy, derivating by the hand the second Piola Kirchhoff tensor (S) and its Jacobian. It was replaced in SOniCS by only defining the strain energy of the desired material model in UFL. The derivative of the strain energy will then be automatically calculated using the ffcx module. Finally, both plugins share the same Forcefield methods for assembling the global residual vector (R) and stiffness matrix (K).}
	\label{fig:pipeline}
\end{figure}

In this section, we present more in-depth the SOniCS (SOFA + FEniCS) plugin. We first introduce the procedure for defining the material model, the element geometry, and the quadrature rule or degree using the UFL (Python) syntax. For simplicity, we only focused on a Saint Venant-Kirchhoff model. But the method can be generalized to all element types following the pipeline shown in figure \ref{fig:pipeline}. Secondly, we explain the methodology for converting the UFL script into efficient C kernels. Finally, we show the interface between the SOniCS plugin and SOFA, stating the conceptual and coding differences. For simplicity and as the modular mechanics plugin's name (Caribou) might change, we use the name SOFA to denote the combination of SOFA and the modular mechanics plugin.  

\subsection{UFL: from FE model to Python code}
The first step is to define the FE model using the UFL syntax in a Python script. As an example, in listing \ref{code}, we describe the 3D simplest hyperelastic model and element: Saint Venant-Kirchhoff with linear Lagrange finite elements on tetrahedra. In the context of hyperelastic simulations, we will exclusively describe features that users could be interested in customizing.

\begin{lstlisting}[language=Python, caption=Python code example (\lstinline{material.py}) of a Saint Venant-Kirchhoff material model using Lagrange linear tetrahedron., label={code}]
# material.py
from ufl import (Coefficient, Constant, Identity,
TestFunction, TrialFunction, inner, ds,
VectorElement, derivative, dx, grad,
tetrahedron, tr, variable)

# Function spaces
cell = tetrahedron
d = cell.geometric_dimension()
element = VectorElement("Lagrange", cell, 1)

# Trial and test functions
du = TrialFunction(element)  # Incremental displacement
v = TestFunction(element)  # Test function

# Functions
u = Coefficient(element)  # Displacement from previous iteration
B = Coefficient(element)   # Body forces
B = Coefficient(element)   # Traction forces

# Kinematics
I = Identity(d)  # Identity tensor
F = variable(I + grad(u))  # Deformation gradient
C = variable(F.T * F)  # Right Cauchy-Green tensor
E = variable(0.5 * (C - I))  # Green-Lagrange tensor

# Elasticity parameters
young = Constant(cell)
poisson = Constant(cell)
mu = young / (2 * (1 + poisson))
lmbda = young * poisson / ((1 + poisson) * (1 - 2 * poisson))

# Stored strain energy density (compressible Neo-Hookean model)
psi = (lmbda / 2) * tr(E) ** 2 + mu * tr(E * E)

# Total potential energy
Pi = psi * dx(degree=1) - inner(B, u) * dx(degree=1) - inner(T, u) * ds(degree=1)

# First variation of Pi (directional derivative about u in the direction of v)
F = derivative(Pi, u, v)

# Compute Jacobian of F
J = derivative(F, u, du)

# Export forms
forms = [F, J, Pi]
\end{lstlisting}

\textbf{Element}: After importing the necessary packages, we can define the element geometry. In listing \ref{code}, on line 10, we used a linear Lagrange tetrahedron. The user can easily modify different parameters of the element, such as the geometry, the family type, or the interpolation degree. For example, by only changing line 10 to
\begin{lstlisting}[numbers=none]
element = VectorElement("Serendipity", hexahedron, 2)
\end{lstlisting}
the element is now a quadratic Serendipity hexahedron. Note that \lstinline{VectorElement} creates by default a function space of vector field equal to the spatial dimension. A complete list of the element available in the Basix documentation~\cite{BasixJoss}.  

\textbf{Material model}: By definition, boundary value problems for hyperelastic media can be expressed as minimization problems. For a domain $\Omega \subset R^{3}$, the goal is to find the displacement field $\bm{u}: \Omega \rightarrow R^{3}$ that minimizes the total potential energy $\Pi$. The potential energy is given by
\begin{equation}
\Pi(\bm{u}) = \int_{\Omega} \psi(\bm{u}) \, \text{d}x - \int_{\Omega} \bm{B} \cdot \bm{u} \, \text{d}x - \int_{\partial\Omega} \bm{T} \cdot \bm{u} \, \text{d}s,
\end{equation}
where $\psi$ is the elastic stored energy density, $\bm{B}$ is a body force (per unit reference volume), and $\bm{T}$ is a traction force (per unit reference area) prescribed on a boundary $\partial\Omega$ (of measure $\text{d}s$) of the domain $\Omega$ (of measure $\text{d}x$).

In listing \ref{code}, at line 37, we define the strain energy of a Saint Venant-Kirchhoff material model as:
\begin{equation}
\label{eq:stvk}
\psi = \frac{\lambda}{2} \text{tr}(\bm{E})^{2} + \mu \text{tr}(\bm{E}^{2}),
\end{equation}
where $\lambda$ and $\mu$ are Lamé material constants while $\bm{E}$ is the Green Lagrange strain tensor. Therefore, we observe that tensor $\bm{E}$ is expressed with respect to the displacement $\bm{u}$, which is the unknown displacement field. Moreover, $\lambda$ and $\mu$ are a function of the Young's modulus and of Poisson's ratio that are assumed to be known constants.

If the user would like to change the material model for Neo-Hookean with the strain energy density
\begin{equation}
\psi = \frac{\mu}{2} (I_{C}-3) - \mu \text{ln}(J) + \frac{\lambda}{2}\text{ln}(J)^{2}.
\end{equation}
It would only require in replacing line $31$ with:
\begin{lstlisting}[numbers=none]
psi = (mu / 2) * (Ic - 3) - mu * ln(J) + (lmbda / 2) * (ln(J)) ** 2
\end{lstlisting}
in addition to previously defining the corresponding kinematics variables $J = \text{det}(\bm{F})$ and $I_{C} = \text{tr}(\bm{C})$.

\begin{lstlisting}[numbers=none]
J = det(F)
I_C = tr(C)
\end{lstlisting}

\textbf{Quadrature rule}: In listing \ref{code}, at line 37, when calculating the total potential energy it is also possible to choose the quadrature rule and the degree. In our example, we selected a quadrature degree of 1, triggering by default the Zienkiewicz and Taylor scheme~\cite{Zienkiewicz2014} for tetrahedra. Hence, the user could also choose to use a Gauss-Jacobi quadrature of degree 2~\cite{Ralston2001} by replacing line 37 with:
\begin{lstlisting}[numbers=none]
Pi = psi * dx(degree=2, scheme="Gauss-Jacobi")  - inner(B, u) * dx(degree=2, scheme="Gauss-Jacobi") - inner(T, u) * ds(degree=2, scheme="Gauss-Jacobi")
\end{lstlisting}.

\subsection{FFCx: from Python code to efficient C kernels}
In the particular case of static hyperelastic simulations, we solve the following non-linear system of equation
\begin{equation}
\label{eq:elasticty}
\bm{K}(\bm{u}) \cdot \text{d}\bm{u} = \bm{R}(\bm{u}) - f(\bm{u}).
\end{equation}
The $\bm{R}$ tensor is called the residual vector and is defined as the Gâteaux derivative of the total potential energy $\Pi$ with respect to change in the displacement $\bm{u}$ in direction $\bm{v}$
\begin{equation}
\bm{R} = \frac{\text{d}\Pi(\bm{u} + \epsilon \bm{v})}{\text{d}\epsilon} |_{\epsilon=0}.
\end{equation}
The tensor $\bm{K}$ is the Jacobian (also called stiffness in the context of mechanics) matrix and corresponds to the derivative of $\bm{R}$
\begin{equation}
\bm{K} = \frac{\text{d}\bm{R}(\bm{u} + \epsilon \bm{\text{d}u})}{\text{d}\epsilon} |_{\epsilon=0}.
\end{equation}

Solving the non linear system in equation \ref{eq:elasticty} can be achieved using the Newton-Raphson algorithm that will iteratively solve a set of linear systems, assuming an initial guess $\bm{u}_n$
\begin{equation}
\bm{u}_{n+1} = \bm{u}_n - d\bm{u}.
\end{equation}

The two tensors can be derived symbolically and exported using the UFL syntax with the function \lstinline{derivative}, as shown at lines 40 and 43 in listing \ref{code}. A simple call to \lstinline{ffcx} will create a .c and .h files containing the code for generating the local $\bm{R}$ and $\bm{K}$ tensors. 

\begin{lstlisting}[numbers=none]
$ ffcx material.py
\end{lstlisting}

\subsection{Integration in SOniCS}
In SOFA, the definitions of the residual and stiffness tensors are carried out within a C++ file, \lstinline{HyperelasticForcefield.cpp}. Each material model is in a separate file. So far, only Saint Venant-Kirchhoff and Neo-Hookean models have been implemented. The users can easily access those functionalities through Python wrappers:

\begin{lstlisting}[numbers=none, language=Python, caption=Python definition of a hyperealstic forcefield in SOFA using a Saint Venant-Kirchhoff material model.]
node.addObject("SaintVenantKirchhoffMaterial", young_modulus=E, poisson_ratio=nu)
node.addObject('HyperelasticForcefield')
\end{lstlisting}

The \lstinline{HyperelasticForcefield} contains several functions, but we are particularly focusing on two of them. The \lstinline{addForce} and \lstinline{assemble_stiffness} functions are assembling the global residual and stiffness tensors respectively. Algorithm \ref{alg:SOFA + Caribou} details the \lstinline{addForce} function, while algorithm \ref{alg:SOniCS} presents our reimplementation of the procedure.  We did not detail the \lstinline{assemble_stiffness} as it involves the exact same differences between the two implementations.

\begin{algorithm}[htp]
	\caption{SOFA addForce function. The addForce function is in charge of assembling the global residual vector.}\label{alg:SOFA + Caribou}
	\begin{algorithmic}[1]
		\For {$\texttt{element in elements}$}
		\State $X \gets \texttt{element.positions}$ \Comment{return the current positions of the element}
		\State $R_{\mathrm{global}} \gets 0$ \Comment{zero the global residual vector of dimension (DOFs $\times 3$)}
		\State $R_{\mathrm{local}} \gets 0$ \Comment{zero the local residual vector of dimension (element DOFs $\times 3$)}
		\For {$\texttt{quadrature in quadratures}$}
		\State $\text{det}J \gets \text{det}(\texttt{quadrature.nodes})$ \Comment{return the Jacobian of the quadrature nodes}
		\State $\text{d}N \gets \texttt{quadrature.nodes.shape\_functions\_derivatives}$ \Comment{return the derivatives of the shape functions of the quadrature nodes}
		\State $w \gets \texttt{quadrature.nodes.weights}$ \Comment{return the weights of the quadrature nodes}
		\State $F \gets X^{T} \cdot \text{d}N$
		\State $J \gets \text{det}(F)$
		\State $C \gets F^{T} \cdot F$
		\State $S \gets \text{f}(C, \texttt{MaterialParameters})$ \Comment{return the second Piola-Kirchhoff depending on the material parameters and kinematics tensors}
		\For {$\texttt{i in \texttt{range(0, NumberOfNodesPerElement)}}$} 
		\State $\text{d}x \gets \text{d}N[\texttt{i}]^{T}$
		\State $R_{local}[i] \gets (\text{det}J \cdot w) \cdot F \cdot S \cdot \text{d}x$ \Comment{allocate the result in the local residual vector}
		\EndFor
		\EndFor
		\For {$\texttt{i in \texttt{range(0, NumberOfNodesPerElement)}}$}
		\State $R_{\texttt{global}}[\texttt{global}(i)] \gets R_{\mathrm{global}}[\texttt{global}(i)] - R_{\mathrm{local}}[i]$ \Comment{$i$ indicates the element node index while $\texttt{global}(i)$ denotes the global node index}
		\EndFor
		\EndFor
	\end{algorithmic}
\end{algorithm}

\begin{algorithm}[htp]
	\caption{SOniCS addForce function. The addForce function is in charge of assembling the global residual vector.}\label{alg:SOniCS}
	\begin{algorithmic}[1]
		\For {$\texttt{element in elements}$}
		\State $X \gets \texttt{element.positions}$ \Comment{return the current positions of the element}
		\State $X_{0} \gets \texttt{element.rest\_positions}$ \Comment{return the initial positions of the element}
		\State $B \gets \texttt{gravity}$ \Comment{return the body forces}
		\State $T \gets \texttt{element.forces}$ \Comment{return the traction forces applied on the element}
		\State $u \gets X-X_{0}$
		\State $R_{\texttt{global}} \gets 0$ \Comment{zero the global residual vector of dimension (DOFs $\times 3$)}
		\State $R_{\texttt{local}} \gets 0$ \Comment{zero the local boundary conditions residual vector of dimension (element DOFs $\times 3$)}
		\State $R_{\texttt{local}}^{\texttt{bc}} \gets 0$ \Comment{zero the local residual vector of dimension (element DOFs $\times 3$)}
		\State $\texttt{constants} \gets \texttt{MaterialParameters}$ \Comment{return the material parameters}
		\State $R_{\texttt{local}} \gets \texttt{tabulate\_tensor}(R_{\texttt{local}}, u, B, \texttt{constants}, X_{0})$
		\State $R_{\texttt{local}}^{\texttt{bc}} \gets \texttt{tabulate\_tensor}(R_{\texttt{local}}, u, T, \texttt{constants}, X_{0})$
		\For {$\texttt{i in \texttt{range(0, NumberOfNodesPerElement)}}$}
		\State $R_{\texttt{global}}[\texttt{\texttt{global}}(i)] \gets R_{\texttt{global}}[\texttt{global}(i)] - (R_{\texttt{local}}[i] + R_{\texttt{local}}^{\texttt{bc}}[i])$ \Comment{$i$ indicates the element node index while $\texttt{global}(i)$ denotes the global node index}
		\EndFor
		\EndFor
	\end{algorithmic}
\end{algorithm}

The structure is similar but we can still observe a few differences. 
\begin{itemize}
	\item The new implementation of algorithm \ref{alg:SOniCS} is more concise and involves less visible tensorial operations because all those operations are efficiently hard coded in the C file provided by FFCx. For example, in algorithm \ref{alg:SOFA + Caribou}, lines 5 to 17 were replaced in the new algorithm \ref{alg:SOniCS} by solely line 11.
	\item Algorithm \ref{alg:SOniCS} needs to have access to the initial position of the object and to the displacement vector. This was indeed not needed in the previous implementation since the modular mechanics plugin takes advantage of writing the deformation gradient only based on the current nodal coordinates $\bm{x}$:
	\begin{equation}
	\bm{F} = \bm{I} + \nabla_{\Omega_0}\bm{u} = \bm{I} + \nabla_{\Omega_0} (\bm{x} - \bm{x_0}) = \nabla_{\Omega_0} \bm{x}.
	\end{equation}
	$\nabla_{\Omega_0}$ and $\bm{x_0}$ respectively denote the gradient and the nodal coordinates in the initial configuration, thus saving one extra vector operation.
	
	\item In the SOFA implementation, the boundary conditions and body forces are treated in separate files. In the new implementation of the Forcefield \ref{alg:SOFA + Caribou}, the boundary conditions and body forces are now directly carried out in the Forcefield on lines 11 and 12. It avoids calling another function to loop again through every element of the object, thus speeding up the assembly of the residual vector.
\end{itemize}

Based on this new implementation, we created a new forcefield \lstinline{HyperelasticForcefield_FEniCS} as close as possible to the existing syntax of \lstinline{HyperelasticForcefield}. We also needed to tune the existing material definition to replace unnecessary calculations and allow us to read the corresponding .c file.

\begin{lstlisting}[numbers=none, language=Python, caption=Python definition of a hyperelastic forcefield in SOniCS using a Saint Venant-Kirchhoff material model.]
node.addObject("FEniCS_Material", material="SaintVenantKirchhoff", young_modulus=E, poisson_ratio=nu)
node.addObject('HyperelasticForcefield_FEniCS')
\end{lstlisting}

\begin{figure}[htp]
	\centering
	\begin{subfigure}{.5\textwidth}
		\centering
		\includegraphics[width=.5\linewidth]{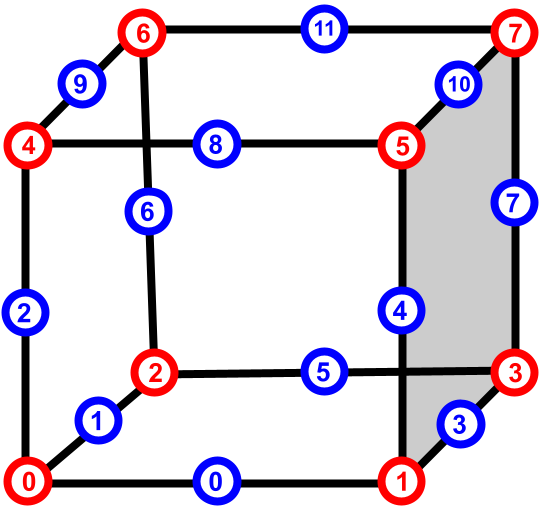}
		\caption{FFCx/Basix.}
		\label{fig:element_fenics}
	\end{subfigure}%
	\begin{subfigure}{.5\textwidth}
		\centering
		\includegraphics[width=.5\linewidth]{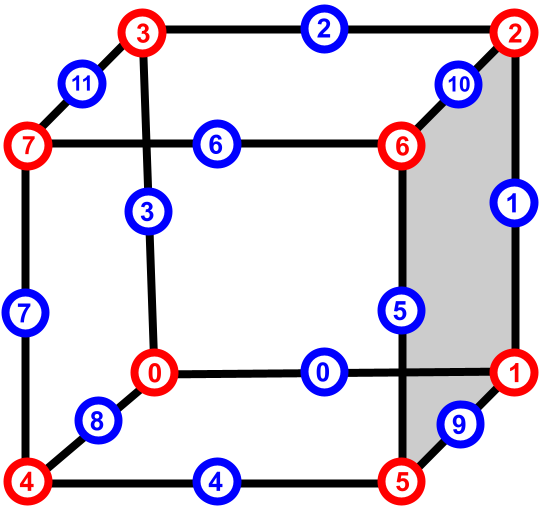}
		\caption{SOFA.}
		\label{fig:element_sofa}
	\end{subfigure}
	\caption{Local numbering of element vertices and edges in both FEniCS and SOFA}
	\label{fig:element_definition}
\end{figure}

Finally, the last hurdle was the element definitions. Indeed, SOFA and FEniCS do not use the same vertices, edges, and facets ordering (as shown in figure \ref{fig:element_definition}). To avoid any conflict with the existing users of SOFA and solve the ordering issue, we proposed rearranging the topology indices, edges, and vertices and create new elements. It ensured an accurate integration over the elements (especially for quadratic Serendipity integrating over the edges) and preserved an appropriate visualization. Those elements have been interfaced with the existing topology named \lstinline{CaribouTopology}.

\begin{lstlisting}[numbers=none, language=Python, caption=Python definition of hexahedron topology in SOFA and SOniCS.]
node.addObject('CaribouTopology', name='topology', template="Hexahedron", indices=mesh.cells_dict['hexahedron'])

node.addObject('CaribouTopology', name='topology', template="Hexahedron_FEniCS", indices=mesh.cells_dict['hexahedron'][:, [4, 5, 0, 1, 7, 6, 3, 2]])
\end{lstlisting}

\section{Numerical examples}
\label{sec:analysis}
In this section we describe three numerical examples used for the validation of our SOniCS implementation.
We use the same domain description for each example while varying the boundary conditions and material parameters of each simulation. 

\begin{figure}
	\centering
	\begin{subfigure}{0.4\textwidth}
		\includegraphics[width=\textwidth]{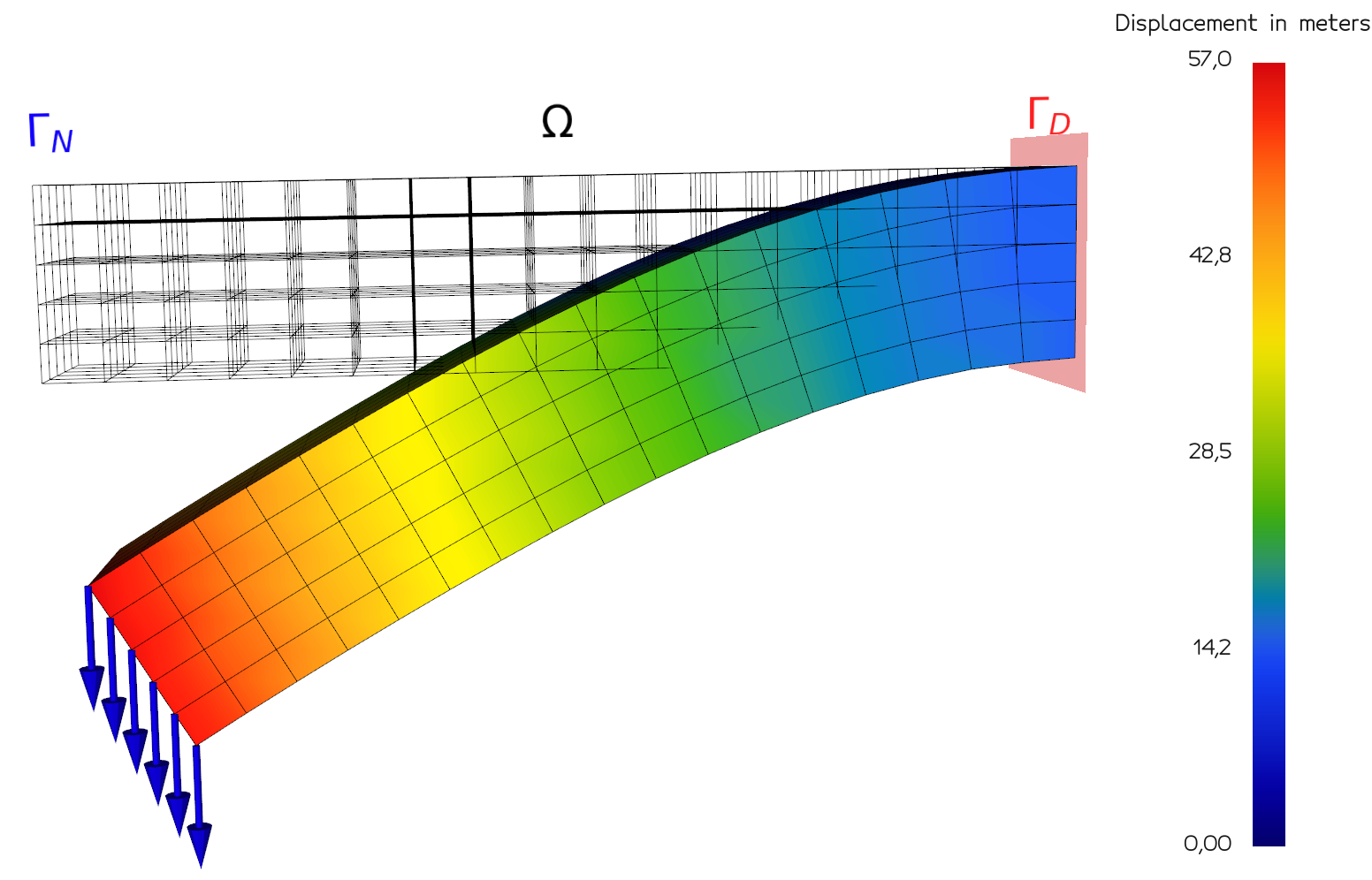}
		\caption{Hexahedral elements.}
		\label{fig:first}
	\end{subfigure}
	\hfill
	\begin{subfigure}{0.4\textwidth}
		\includegraphics[width=\textwidth]{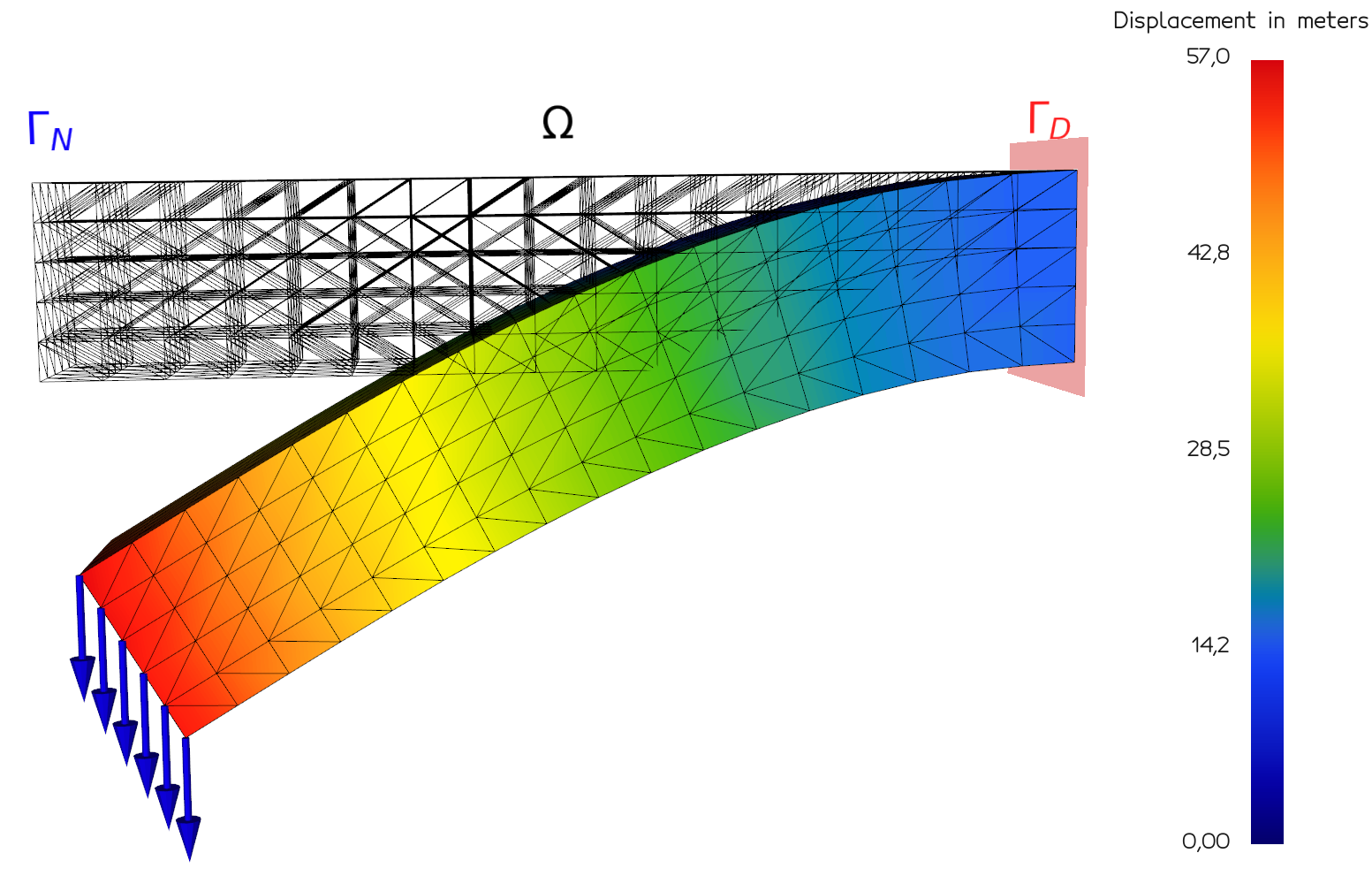}
		\caption{Tetrahedral elements.}
		\label{fig:second}
	\end{subfigure}
	
	\caption{Cantilever beam domain discretization and displacement field. $\Omega$ is a domain represented by a squared-section beam of dimensions $80 \times 15 \times 15$\si{m^3}, considered fixed on the right side ($\bm{u}=0$ on $\Gamma_D$) while Neumann boundary conditions are applied on the left side ($\Gamma_N$).}
	\label{fig:figures}
\end{figure}

Let $\Omega$ be a domain represented by a squared-section beam of dimensions $80 \times 15 \times 15$\si{m^3}, considered fixed on the right side ($\bm{u}=0$ on $\Gamma_D$) while Neumann boundary conditions are applied on the left side ($\Gamma_N$), as shown in figure \ref{fig:first} and \ref{fig:second}.

$\Omega$ was discretized using two different geometrical elements using linear and quadratic interpolations. P1 and P2 elements stand for linear or quadratic tetrahedra, while Q1 and Q2 denote linear and quadratic hexahedra.

To solve each hyperlastic formulation described in equation \ref{eq:elasticty} we used an identical implementation of the classical Newton Raphson (NR) solver for SOniCS, SOFA and FEBio. The solver had the following parameters: a maximum of 25 iterations with a residual and displacement tolerance of $10^{-10}$. In order to compare the running time of the two implementations, we, therefore, introduced the mean NR iteration time. We defined the NR iteration time as the duration for assembling and factorizing the system matrix, solving and propagating the unknown increment, updating, and computing the force and displacement residual. After checking that the same number of iterations have been achieved, we averaged the total time over the number of iterations needed for the solver convergence. All calculations were performed using an Intel® Core™ i5-6300HQ CPU @ 2.30GHz × 4 processor with a 16GiB memory and a NV117 / Mesa Intel® HD Graphics 530 (SKL GT2) graphics card.

We evaluated the soundness of the SOniCS solution using SOFA, FEBio or a manufactured solution as the reference solution and computed the Euclidean relative $L^{2}$ error ($e(\bm{u},\bm{v})$)~\cite{odot2022}.

\begin{equation}
e(\bm{u},\bm{v}) = \frac{ || \bm{u} - \bm{v} || }{|| \bm{v} ||},
\label{Emean}
\end{equation}
where $\bm{u}$ and $\bm{v}$ are the calculated displacements for SOniCS and the reference implementation, respectively.

In this section, we first compare our solution with an analytical one: the manufactured solution. Then, we consider a clamped cantilever beam subject to Neumann boundary conditions and compare its deformation with the SOFA solution. Finally, using the same cantilever beam, we implemented a Mooney Rivlin model (uncoded in SOFA) using SOniCS and compared the solution with FEBio.

\subsection{Manufactured solution}
\label{sec:manufactured}
Aiming at code verification, the method of the manufactured solution consists in choosing an exact solution to the problem as an analytical expression~\cite{Chamberland2010}. The chosen analytical expression is then inserted into the Partial Differential Equation (PDE) under consideration to find the conditions that lead to this solution. In general, the manufactured chosen solution is expressed in simple primitive functions like $\text{sin(), exp(), tanh()}$, etc... In the context of hyperelastic equations, we considered the following manufactured solution for the displacement

\begin{equation}
\label{manufactured_displacements}
\bm{u}(x, y, z) = \begin{bmatrix}
10^{-2}\cdot z \cdot e^{x}\\
10^{-2}\cdot z \cdot e^{y}\\
10^{-2}\cdot z \cdot e^{z}
\end{bmatrix} \text{on} \, \Omega.
\end{equation}

Starting from the above chosen displacement and using continuum mechanics laws, the relative analytical forces are applied as Neumann boundary conditions and deduced as follows

\begin{equation}
\label{eq:deformation_gradient}
\bm{F} = \bm{I_d} + \text{grad}(\bm{u}),
\end{equation}
\begin{equation}
\bm{P} = \frac{\partial W}{\partial \bm{F}},
\end{equation}
\begin{equation}
\bm{f} = - \mathbf{\nabla} \cdot \bm{P} \, \text{on} \, \Gamma_{N}.
\end{equation}

Where $\bm{F}$ is the deformation gradient, $\bm{I_d}$ is the identity matrix of dimension $d$, $\bm{P}$ is the first Piola-Kirchhoff stress tensor and $W$ is the strain energy density depending on the material model constitutive law. We used a Saint Venant-Kirchhoff material \eqref{eq:stvk} with a Young's Modulus of \SI{3}{kPa} and a Poisson's ratio of $0.3$, the computation of this solution was performed using Python Sympy package~\cite{sympy}.

\begin{figure}[ht]
	\centering
	\includegraphics[width=0.75\linewidth]{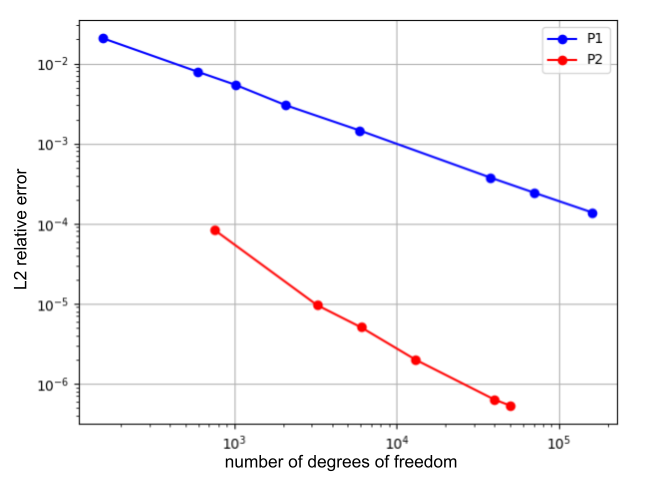}
	\caption{Plot of the mesh convergence analysis of the manufactured solution. The $L^2$ error between the analytical and the SOniCS simulation is calculated for different mesh refinement with fixed parameters ($E$=\SI{3}{kPa} and $\nu$=0.3) for P1 linear tetrahedra (blue) and P2 quadratic tetrahedra (red) elements.}
	\label{fig:loglog}
\end{figure}

In this experiment, we generated $8$ and $6$ discretizations of P1 and P2 elements, respectively, with a decreasing element size in both scenarios. For each discretization, we applied the relative analytical forces deduced from the manufactured solution \eqref{manufactured_displacements} and used SOniCS Saint Venant-Kirchhoff material implementation with the same parameters as Sympy’s to fill the domain. The displacements obtained were compared to the chosen analytical solution in equation \eqref{manufactured_displacements} for each discretization. The results are presented in figure \ref{fig:loglog}, the error metric is the relative mean error in equation \eqref{Emean}.

\subsection{Benchmark with SOFA}
\label{section:beam}
The cantilever beam deflection is a classical mechanical test case, as you can smoothly refine the mesh due to the simplicity of the geometry or modify its boundary conditions to fit real-life experiments. In this context, the beam was still clamped on the right side (natural Dirichlet condition on $\Gamma_D$) while Neumann boundary conditions were applied on the left side ($\Gamma_N$). To compare our SOniCS implementation, we model the deformation of the beam with two hyperelastic material models: Saint Venant-Kirchhoff and Neo-Hookean. We fixed the mechanical parameters and the Neumann boundary conditions equal to \SI{-10}{Pa} in the $y$ direction, until reaching sufficient large deformations with the same parameters as before: $E = \SI{3}{kPa}$ and $\nu = 0.3$.

This study aims at comparing the finite element solutions provided by SOniCS and SOFA under the same constraints in terms of computational and running time performances. To do so, we computed the mean relative $L^2$ error ($e(\bm{u_{\texttt{SOniCS}}},\bm{u_{\texttt{SOFA}}})$) between the SOniCS solution using SOFA as the reference solution.

The results obtained are presented in tables \ref{SOniCSSOFA + CaribouStVK} and \ref{SOniCSSOFA + CaribouNH} for Saint Venant-Kirchhoff and Neo-Hookean materials, respectively.

\begin{table}[!htbp]
	\centering
	\begin{tabular}{ |p{2cm}|p{2cm}|p{3cm}|p{3cm}|p{2cm}|  }
		\hline
		\multicolumn{5}{|c|}{Saint Venant-Kirchhoff material model} \\
		\hline
		Element& Number of DOFs & SOniCS mean NR iteration time (\si{ms})& SOFA mean NR iteration time (\si{ms}) &$e$\\
		\hline
		P1 & 351 & 2.00 & 4.4 & 5.87e-14 \\
		P2 & 1875 & 8.16 & 13.33 & 7.14e-14 \\
		Q1 & 351 & 2.00 & 7.16 & 1.53e-13 \\
		Q2  & 1143 & 15.66 & 16.33 & 1.66e-13 \\
		\hline
	\end{tabular}
	\caption{\label{SOniCSSOFA + CaribouStVK} Mean relative error ($e(\bm{u_{\texttt{SOniCS}}},\bm{u_{\texttt{SOFA}}})$ defined in equation \ref{Emean}) and mean NR (Newton-Raphson) iteration time between SOniCS and SOFA for different element geometries and interpolation schemes using Saint Venant-Kirchhoff material model. P1 and P2 elements stand for linear or quadratic tetrahedra, while Q1 and Q2 denote linear and quadratic hexahedra.}
\end{table}

\begin{table}[!htbp]
	\centering
	\begin{tabular}{ |p{2cm}|p{2cm}|p{3cm}|p{3cm}|p{2cm}|  }
		\hline
		\multicolumn{5}{|c|}{Neo-Hookean material model} \\
		\hline
		Element& Number of DOFs & SOniCS mean NR iteration time (\si{ms})& SOFA mean NR iteration time (\si{ms}) &$e$\\
		\hline
		P1 & 351 & 1.17 & 1.8 & 2.17e-14 \\
		P2 & 1875 & 21.16 & 29.16 & 7.53e-14 \\
		Q1 & 351 & 2.16 & 7.3 & 1.89e-13 \\
		Q2  & 1143 & 17.14 & 20.42 & 1.64e-13 \\
		\hline
	\end{tabular}
	\caption{\label{SOniCSSOFA + CaribouNH} Mean relative error ($e(\bm{u_{\texttt{SOniCS}}},\bm{u_{\texttt{SOFA}}})$ defined in equation \ref{Emean}) and mean NR (Newton-Raphson) iteration time between SOniCS and SOFA for different element geometries and interpolation schemes using Neo-Hookean material model. P1 and P2 elements stand for linear or quadratic tetrahedra, while Q1 and Q2 denote linear and quadratic hexahedra.}
\end{table}

\subsection{Benchmark with FEBio}
\label{sec:febio}
FEBio is a open-source finite element package specifically designed for biomechanical applications. It offers modeling scenarios, a wide range of constitutive material models, and boundary conditions relevant to numerous research areas in biomechanics. In this section, FEBio was used to compute the same scenarios as in section  \ref{section:beam} to evaluate the trustworthiness of SOniCS. A more advanced constitutive material model, Mooney Rivlin, was introduced for this purpose

\begin{equation}
\psi=C_{01}\left(\overline{\mathrm{I}_{C}}-3\right)+C_{10}\left(\overline{\mathrm{I}\mathrm{I}_{C}}-3\right)+\frac{K}{2}\text{ln}(J).
\end{equation}
Where $C_{01}$, $C_{10}$, and $K$ are the material constants in addition to the modified invariants $\overline{\mathrm{I}_{C}} = J^{-\frac{2}{3}} \; \mathrm{I}_{C}$, $\overline{\mathrm{II}_{C}} = J^{-\frac{4}{3}} \; \mathrm{II}_{C}$ defined based on the classic invariants $\mathrm{I}_{\mathrm{\boldsymbol{C}}} =\operatorname{tr}(\mathbf{C})$, 
$\mathrm{II}_{\mathrm{\boldsymbol{C}}} =\frac{1}{2}\left((\operatorname{tr}(\mathbf{C}))^{2}-\operatorname{tr}\left(\mathbf{C}^{2}\right)\right)$. In order to obtain sufficiently large deformations, we chose the following material parameters: $C_{01} = \SI{2000}{Pa}$, $C_{10} = \SI{100}{Pa}$, and $K = \SI{1000}{Pa}$.

Tables \ref{SOniCSFEBioStVK}, \ref{SOniCSFEBioNH}, \ref{SOniCSFEBioMR} show the results obtained for Saint Venant-Kirchhoff, Neo-Hookean and Mooney Rivlin material models and considering the four discretizations implemented so far in SOniCS. The error evaluation is still based on the mean relative error defined in equation \ref{Emean} using FEBio as the reference while using a Newton Raphson solver with the same characteristics in both cases.
\\
\begin{table}[ht]
	\centering
	\begin{tabular}{ |p{2cm}|p{2cm}|p{3cm}|p{3cm}|p{2cm}|  }
		\hline
		\multicolumn{5}{|c|}{Saint Venant-Kirchhoff material model} \\
		\hline
		Element& Number of DOFs & SOniCS mean NR iteration time (\si{ms})& FEBio mean NR iteration time (\si{ms}) &$e$\\
		\hline
		P1 & 351 & 1.6 & 9.11 & 6.01e-10 \\
		P2 & 1875 & 18.25 & 20.95 & 0.08 \\
		Q1 & 351 & 4.60 & 5.66 & 4.19e-10 \\
		Q2  & 1143 & 16.75 & 30.7 & 0.13 \\
		\hline
	\end{tabular}
	\caption{\label{SOniCSFEBioStVK}Mean relative error ($e(\bm{u_{\texttt{SOniCS}}},\bm{u_{\texttt{FEBio}}})$ defined in equation \ref{Emean}) and mean NR (Newton-Raphson) iteration time between SOniCS and FEBio for different element geometries and interpolation schemes using Saint Venant-Kirchhoff material model. P1 and P2 elements stand for linear or quadratic tetrahedra, while Q1 and Q2 denote linear and quadratic hexahedra.}
\end{table}

\begin{table}[ht]
	\centering
	\begin{tabular}{ |p{2cm}|p{2cm}|p{3cm}|p{3cm}|p{2cm}|  }
		\hline
		\multicolumn{5}{|c|}{Neo-Hookean material model} \\
		\hline
		Element& Number of DOFs & SOniCS mean NR iteration time (\si{ms})& FEBio mean NR iteration time (\si{ms}) &$e$ \\
		\hline
		P1 & 351 & 5.20 & 8.78 & 6.53e-10 \\
		P2 & 1875 & 14.30 & 19.10 & 1.5e-2 \\
		Q1 & 351 & 4.83 & 5.41 & 8.08e-10 \\
		Q2  & 1143 & 18.00 & 28.10 & 0.09 \\
		\hline
	\end{tabular}
	\caption{\label{SOniCSFEBioNH}Mean relative error ($e(\bm{u_{\texttt{SOniCS}}},\bm{u_{\texttt{FEBio}}})$ defined in equation \ref{Emean}) and mean NR (Newton-Raphson) iteration time between SOniCS and FEBio for different element geometries and interpolation schemes using Neo-Hookean material model. P1 and P2 elements stand for linear or quadratic tetrahedra, while Q1 and Q2 denote linear and quadratic hexahedra.}
\end{table}

\begin{table}[ht]
	\centering
	\begin{tabular}{ |p{2cm}|p{2cm}|p{3cm}|p{3cm}|p{2cm}|  }
		\hline
		\multicolumn{5}{|c|}{Mooney Rivlin material model} \\
		\hline
		Element& Number of DOFs & SOniCS mean NR iteration time (\si{ms})& FEBio mean NR iteration time (\si{ms}) &$e$ \\
		\hline
		P1 & 351 & 3.75 & 8.53 & 2.49e-9 \\
		P2 & 1875 & 15.81 & 16.85 & 9.97e-3 \\
		Q1 & 351 & 4.00 & 5.76 & 10.92 \\
		Q2  & 1143 & 22.28 & 23.61 & 4.23e-2 \\
		\hline
	\end{tabular}
	\caption{\label{SOniCSFEBioMR}Mean relative error ($e(\bm{u_{\texttt{SOniCS}}},\bm{u_{\texttt{FEBio}}})$ defined in equation \ref{Emean}) and mean NR (Newton-Raphson) iteration time between SOniCS and FEBio for different element geometries and interpolation schemes using Mooney Rivlin material model. P1 and P2 elements stand for linear or quadratic tetrahedra, while Q1 and Q2 denote linear and quadratic hexahedra.}
\end{table}

\section{Hozapfel and Ogden anisotropic material model coupled with haptic simulation}
\label{sec:haptic}
In the context of numerical surgical simulations, a robot haptic feedback has been shown to be a consistent tool for drastically improving user-interactions and opening up countless applications. Among them, haptic devices have mainly been used as a training tool for surgeons. Indeed prior to surgery, under the assumption of known geometry and mechanical properties of the patient's organ, a surgeon would be able to plan and better choose between specific surgical paths/approaches. In this paper, we used the 3D Systems Touch Haptic Device robot coupled with the SOFA plugin Geomagic to allow interactions between the instrument and the simulations. 
For this hypothetical simulation, we virtually simulate the contact between a surgical tool and a liver during surgery. The liver was described by an anisotropic Holzapfel Ogden model~\cite{Holzapfel2009, Pezzuto2014}, with an existing FEniCS implementation~\cite{Hauseux2018}.

\begin{equation}
\psi = \psi_{\text {iso }} (\bm{F}) + \psi_{\text {vol}}(J).
\end{equation}

$\psi_{\text {iso }}$ and $\psi_{\text {vol}}$ are the isochoric and volumetric part of the strain energy density function respectively. The volumetric part can be evaluated as a function of the bulk modulus $\kappa$ of the material and $J$

\begin{equation}
\psi_{\text {vol }} (J) = \frac{\kappa}{4} (J^{2} - 1 - 2 \text{ln}(J)),
\end{equation}
\begin{equation}
\begin{aligned}
\psi_{\text {iso }} (\bm{F})=\frac{a}{2 b} \exp \left[b\left(\mathrm{I}_{1}-3\right)\right]+
& \sum_{i=f, s} \frac{a_{i}}{2 b_{i}} \exp \left[b_{i}\left(\mathrm{I}_{4 i}-1\right)^{2}\right] + \\
&\frac{a_{\mathrm{fs}}}{2 b_{\mathrm{fs}}}\left(\exp \left[b_{\mathrm{fs}} \mathrm{I}_{8 \mathrm{fs}}^{2}\right]-1\right),
\end{aligned}
\end{equation}

with
\begin{equation}
\begin{split}
\mathrm{I}_{4f} = \bm{f_{0}} \cdot \bm{C} \cdot \bm{f_{0}}, \ \mathrm{I}_{4s} = \bm{s_{0}} \cdot \bm{C} \cdot \bm{s_{0}} ,\ \mathrm{and} \ \mathrm{I}_{8 \mathrm{fs}} = \bm{f_{0}} \cdot \bm{C} \cdot \bm{s_{0}}.
\end{split}
\end{equation}

The transversely isotropic behavior can be obtained by removing the parameters $a_{fs}$, $b_{fs}$, $a_s$ and $b_s$, while the isotropic behavior is obtained by also suppressing the two parameters $a_f$ and $b_f$. This kind of model is frequently used to model orthotropic materials (e.g. muscle with fibers or tendons). Vectors $\bm{f_0}$, $\bm{s_0}$ are the unit base vectors normal to the planes of symmetry. For our application, we selected the same material properties as chosen in~\cite{Hauseux2018}.

The instrument is assumed to be a rigid body kinematically constrained by the haptic device position at each time step. The contact forces generated by the collision between the instrument and the liver are calculated using frictional contact and an implicit Euler scheme to solve the dynamic system~\cite{Courtecuisse2013}. Finally, the contact forces are transmitted back to the user's hand through the haptic device. The result is a simulation running at $100$ FPS (Frames Per Seconds) on average displayed in figure \ref{fig:haptic}. A full video of the interaction with the haptic device is available in the Supplementary Materials. The real-time performance has been obtained by using a small number of DOFs ($543$). A higher number of DOFs could be used once a GPU implementation is available. 

\begin{figure}[ht]
	\centering
	\includegraphics[width=0.6\linewidth]{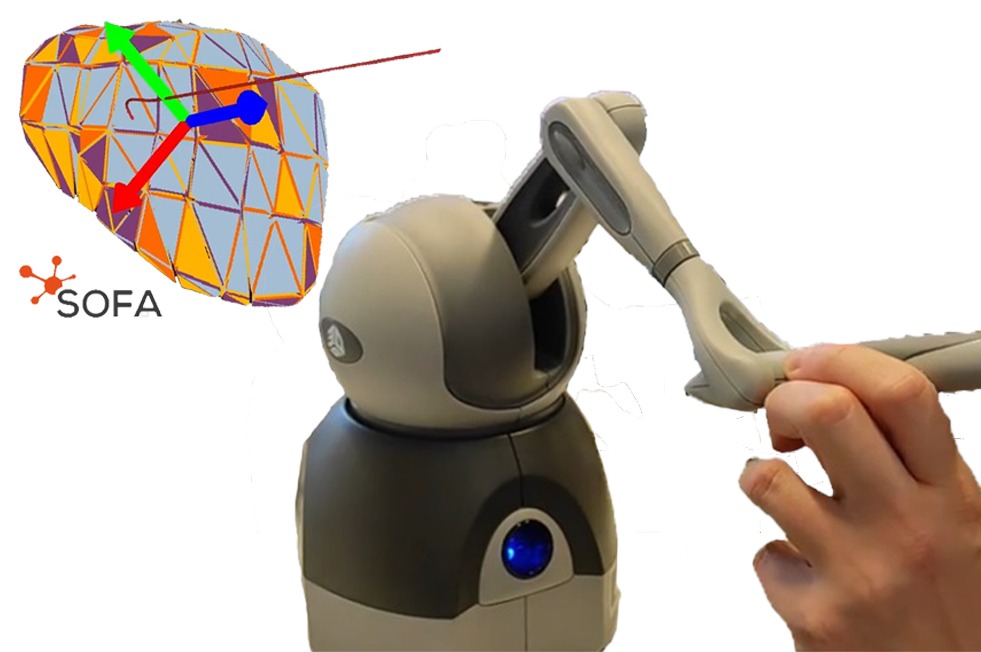}
	\caption{Numerical simulation of a liver in contact with a surgical tool connected to a haptic device. On the right, the 3D Systems Touch Haptic Device is controlled by the user. On the left, the liver is modeled using a Holzapfel Ogden anisotropic material in contact with the surgical tool (in red) guided by the user. In case of contact detection, the contact forces are transmitted to the user through the haptic device. A video of the simulation is available as supplementary materials.}
	\label{fig:haptic}
\end{figure}

\section{Discussion}
We shows that the SOniCS plugin is an efficient implementation of material models for hyperelastic simulations and enables the user to develop an intuitive understanding of the impact of modelling choices on accuracy reliability of the predictions. We first demonstrated a convergence study for Saint Venant-Kirchhoff material using P1 and P2 elements with the manufactured solution in section \ref{sec:manufactured}. Indeed, as expected, by refining the mesh, the $L^2$ relative error almost linearly decreases when increasing the number of DOFs (on a log-log plot), showing the stability of our method.

Then, from section \ref{section:beam}, two main results are noteworthy from tables \ref{SOniCSSOFA + CaribouStVK} and \ref{SOniCSSOFA + CaribouNH}. First, the relative error of the displacement between SOniCS and SOFA, for both material models, is close to machine precision for P1, P2, Q1, and Q2 discretizations. Finally, the last comment concerns the mean NR iteration time. We observe that the SOniCS implementation is slightly faster than SOFA. The difference increases when using more DOFs, e.g., \SI{2.40}{ms} difference for P1 against \SI{5.17}{ms} difference for P2 elements for the Saint Venant-Kirchhoff model. The reason for this difference is the need for SOFA to compute the shape functions and derivatives, then calculate the local residual and Jacobian using multiple tensor operations. Conversely, SOniCS has all those operations efficiently hard-coded in C kernels, thus performing faster than SOFA. 

We compared the SOniCS and FEBio simulations for several material models: Saint Venant-Kirchhoff, Neo-Hookean, and Mooney-Rivlin in section \ref{sec:febio}. For P1 elements, the errors are close to machine precision for all $3$ models. For P2 and Q2 elements, the $3$ models display similar errors that still represent a minor error (less than $0.08$ and $0.1$, respectively) which is of same magnitude between SOFA and FEBio, as well as FEniCS and FEBio. The reasons for those minor errors could be the difference in the implementation of the elements or in the choice of solver parameters. For Q1 elements, we reached an accuracy near machine precision for Saint Venant-Kirchhoff and Neo-Hookean. However, the relative error rose close to $11$ for the Mooney-Rivlin model. To further understand this divergence, we used a third open-source software AceGen. AceGen also uses an automatic code generation package for the symbolic generation of new finite elements. The cantilever beam scenario presented in section \ref{sec:febio} was reproduced using AceGen under the same conditions and with a Q1 discretization of the domain. Using the same metric as defined in equation \ref{Emean}, the results are the following: $e(\bm{u_{\texttt{SOniCS}}},\bm{u_{\texttt{Acegen}}}) = 3.33 $ and $e(\bm{u_{\texttt{FEBio}}},\bm{u_{\texttt{Acegen}}}) = 14.19 $. Even if SOniCS and Acegen showed similar results, a more in-depth study would be needed to confirm the soundness of our solution. Despite using a finer mesh, the error was getting slightly lower but was still noticeable. Eventually, FEBio has shown difficulty in converging with trivial parameter sets or when increasing the number of DOFs. Thus, as shown in tables \ref{SOniCSFEBioStVK},\ref{SOniCSFEBioNH}, and \ref{SOniCSFEBioMR}, on average, SOniCS is solving the equation system faster than FEBio. Several reasons could explain those differences, such as the number of quadrature points used, the implementation of the Newton-Raphson scheme, or the solver parameters.

Finally, we showed the capabilities of the SOniCS plugin in simulating complex material models such as the Holzapfel Ogden anisotropic model coupled with a haptic device in section \ref{sec:haptic}. The material model was effortlessly implemented for several elements (P1, P2, Q1, and Q2) without needing any manual derivation or coding. The result is a real-time simulator functional for surgeons' training or any other biomechanics simulation replicating the behavior of a liver in contact with a surgical tool. It was not possible to verify our simulation as we did not find any similar implementation of such material in any open-source software.

\section{Conclusion and outlook}
We performed several numerical experiments to develop intuitive understanding of new material models in SOFA using the SOniCS plugin. First, we validated the most common hyperelastic material models: Saint Venant-Kirchhoff and Neo-Hookean using a manufactured solution. Then, utilizing FEBio as a reference, we validated our implementation of a Mooney-Rivlin material model using P1, P2, Q1, and Q2 elements. The final application employed a haptic device to interact with an anisotropic Holzapfel Ogden liver model in real-time.

The study used our SOniCS plugin to generate optimized C code for complex material models compatible with SOFA. On one side, we benefited from FEniCS automatic differentiation and code generation capabilities to bypass the difficulties of deriving and implementing the consistent Jacobian in SOFA. On the user side, we implemented compatible and user-friendly SOFA forcefields to use the FEniCS C kernels. A SOFA user can now easily define a new material model by specifying its strain energy function, element geometry or family and the quadrature scheme and degree only in Python. We made the open-source code and all data and test cases available as the supplementary material.

In future work, we intend to apply the SOniCS plugin for solving more complex mechanical phenomena. For example, mixed formulations for solving incompressible materials, viscous or plastic effects, and multi-material systems. In this paper, we only utilized the plugin for solving hyperelasticity equations, but it would be interesting to tackle multi-physics problems such as thermomechanics or magnetomechanics. We only used Lagrangian P1, P2, Q1, and Q2 elements while more geometries such as prisms could be relevant for the SOFA and FEniCS community. Finally, our validation only focused on the deformation field of the structures, while a more in-depth study would also enable us to validate an analytical stress field.

Some work remains to improve user-friendliness of our package. Indeed, even if writing the Python file describing the material model and generating the associated C file is mostly effortless and automated, some steps are still manual. Indeed, the \lstinline{FEniCS_Material} C++ class must have knowledge of every new C file created at compile time. Hence, for the moment, the user still has to manually specify two C++ functions in the code and recompile the whole plugin. Even if this step is manageable, it still requires diving into the C++ code, which can discourage a few users. Meanwhile, to mitigate this effect, actions have been taken by providing detailed documentation and tutorials for this crucial step. In addition, future works aim at improving this stage by directly providing the path of the C file in the Python code to trigger just-in-time compilation for new materials.

\section*{Acknowledgements}
This project has received funding from the European Union’s Horizon 2020 research and innovation programme under the Marie Sklodowska-Curie grant agreement No. 764644. This abstract only contains the RAINBOW consortium’s views and the Research Executive Agency and the Commission are not responsible for any use that may be made of the information it contains. This publication has been prepared in the framework of the DRIVEN project funded by the European Union’s Horizon 2020 Research and Innovation programme under Grant Agreement No. 811099. This study is supported by the National Research Fund, Luxembourg, and cofunded under the Marie Curie Actions of the European Commission (FP7-COFUND) Grant No. 6693582. This project is supported by the "IHU Strasbourg - Institut de chirurgie guidée par l'image" Strasbourg, France. The authors would like to acknowledge Thomas Lavigne for providing the Acegen simulation, editing Figure 5, and reviewing the manuscript. The authors would also like to thank Dr. Michal Habera and Dr. Matthew Scroggs for their help with the quadratic hexahedron Serendipity elements.

\bibliographystyle{apalike}
\bibliography{biblio}

\end{document}